\documentclass{article}
\usepackage{spconf,amsmath,amssymb,graphicx,booktabs,hyperref,tikz}
\usepackage{newtxtext,newtxmath}
\usepackage{multirow}
\hypersetup{hidelinks}

\title{REDUCING THE OUTPUT-MODE GAP IN SPEECH LANGUAGE MODELS VIA JOINT-OUTPUT ON-POLICY DISTILLATION}

\name{Daxin Tan$^{1}$, Dehua Tao$^{1}$, Chengxi Deng$^{2}$, Hanlin Zhang$^{3}$,
Xiao Chen$^{1}$}
\address{
$^{1}$AI Lab, Leibniz Research Center, Huawei\\
$^{2}$The Chinese University of Hong Kong\\
$^{3}$City University of Hong Kong\\
\texttt{\{tan.daxin1, chen.xiao2\}@huawei.com}}

\begin{document}
\ninept
\maketitle

\begin{abstract}
Autoregressive generation of interleaved text and acoustic tokens is a common approach to spoken-response generation in speech large language models. Although this design enables streaming generation with explicit textual guidance, generated acoustic tokens become part of the context for subsequent text predictions. Given identical speech inputs, we observe markedly lower answer accuracy for the internal text generated in speech-to-text-and-speech (S2TS) mode than for speech-to-text (S2T) responses. We term this discrepancy the \emph{output-mode gap} (OMG). To reduce OMG, we propose \emph{Joint-Output On-Policy Distillation} (JO-OPD), which distills the model's stronger S2T policy into joint generation using student-generated S2TS trajectories. At each text position, the S2T teacher provides soft targets from a text-only projection of the student's preceding outputs, while the student predicts from the corresponding full interleaved history. A preservation objective further regularizes native non-text predictions. Experiments on Step-Audio-2-mini and Baichuan-Audio-Instruct reveal OMG across two interleaved generation architectures. On Step-Audio-2-mini, JO-OPD reduces OMG from 42.87 to 16.26 percentage points on Spoken-MQA and from 29.72 to 13.04 points on speech-rendered GSM8K, with little change in S2T accuracy and substantially larger reductions than matched SFT baselines. ASR-based evaluation further shows a 7.49-point improvement in spoken-answer accuracy on Spoken-MQA.
\end{abstract}


\section{Introduction}

Speech provides a natural interface for human--machine communication. Speech large language models (SLLMs) seek to bring the knowledge and reasoning capabilities of language models to spoken interaction. Models such as Qwen-Audio~\cite{chu2023qwenaudio} and SALMONN~\cite{tang2023salmonn} process speech and audio inputs to produce textual responses. Systems including SpeechGPT~\cite{zhang2023speechgpt}, Moshi~\cite{defossez2024moshi}, and GLM-4-Voice~\cite{zeng2024glm4voice} further support spoken responses, extending language-model capabilities from speech understanding to speech generation. To support linguistic control over response content, many speech-output systems retain an explicit text stream to guide acoustic generation.

Two representative approaches differ in how they couple text and speech generation. One approach adopts a Thinker--Talker architecture, as exemplified by Qwen3-Omni~\cite{xu2025qwen3omni}, which separates text generation from speech generation, with a Talker producing speech conditioned on text outputs from a Thinker. An alternative approach uses fully discrete models, such as Step-Audio~2~\cite{wu2025stepaudio2} and Baichuan-Audio~\cite{li2025baichuan}, which represent both text and speech as discrete tokens and generate them autoregressively in an interleaving pattern. This design supports streaming speech generation while incorporating previously generated acoustic tokens into the context for subsequent text predictions. In this interleaved setting, we seek to answer the following question: \emph{given the same speech input, does a model preserve its text answer accuracy when it also generates speech?}

A line of work studies input-side modality gaps, examining how speech-conditioned performance differs from text-input performance and how speech adaptation affects pretrained language-model capabilities~\cite{xiang2025modalitygap,cuervo2025salad}. Efforts to narrow these gaps include cross-modal alignment and distillation, reinforcement learning, and text-based input representations augmented with prosodic information~\cite{cao2026xopd,hu2026cord,wang2026tars,cui2026textpro}. Complementing these input-side studies, we investigate an output-side discrepancy by keeping the speech input and model fixed and comparing the accuracy of text-only responses with that of the internal text generated during joint text--speech generation.

Paired evaluations of Step-Audio-2-mini and Baichuan-Audio-Instruct reveal that, given identical speech inputs, answer accuracy is substantially lower for the internal text generated in speech-to-text-and-speech (S2TS) mode than for speech-to-text (S2T) responses. We term this difference the \emph{output-mode gap} (OMG), measured as S2T accuracy minus S2TS internal-text accuracy under the same answer-scoring rule. On Spoken-MQA and speech-rendered GSM8K, OMG reaches 42.87 and 29.72 percentage points (pp) for Step-Audio-2-mini, and 12.41 and 10.39 points for Baichuan-Audio-Instruct, respectively. Explicit reasoning instructions further widen OMG on human-recorded VoiceBench-BBH questions for Step-Audio-2-mini, extending the observation beyond mathematical tasks. These findings motivate improving text answer accuracy within interleaved text--speech generation.

Motivated by the model's stronger S2T performance, we propose \emph{Joint-Output On-Policy Distillation} (JO-OPD) to reduce OMG using its own S2T policy as a teacher. JO-OPD distills the teacher's text predictions along student-generated S2TS trajectories. With the speech input unchanged, the teacher conditions on the text-only projection of each student prefix, while the student conditions on the full interleaved history. An additional preservation objective regularizes native non-text predictions to limit changes to the model's speech-generation behavior.

Our contributions are threefold. \textbf{(1)} We identify and quantify the output-mode gap in Step-Audio~2 and Baichuan-Audio, observing substantially lower text answer accuracy when speech output is enabled under identical speech inputs. \textbf{(2)} We propose JO-OPD, which distills a model's stronger S2T policy along student-generated S2TS trajectories while regularizing native non-text predictions. \textbf{(3)} Experimental results demonstrate that JO-OPD reduces OMG in both models, and further analyses examine how internal-text improvements translate into spoken-answer accuracy.

\section{Related Work}

\subsection{Coupling text and speech generation}

Spirit~LM~\cite{nguyen2024spiritlm} autoregressively models interleaved text and speech tokens, while MiMo-Audio~\cite{coreteam2025mimoaudio} jointly models text tokens and audio patches. LongCat-Next~\cite{longcat2026} compares text-guidance accuracy under parallel and serial audio generation. TtT~\cite{liu2026ttt} combines autoregressive text generation with non-autoregressive audio diffusion, whereas PRIME-Speech~\cite{hu2026primespeech} adds a trainable audio post-decoder to a frozen speech-to-text backbone. These studies explore generation designs and capability preservation. JO-OPD targets the output-mode discrepancy within an existing interleaved architecture through distillation and non-text prediction regularization.

\subsection{Input-side modality gaps}

Prior work has observed that extending text LLMs into speech LLMs can degrade knowledge and reasoning capabilities~\cite{xiang2025modalitygap,cuervo2025salad}. To address this issue, SALAD~\cite{cuervo2025salad} combines cross-modal distillation with active data selection, while TARS~\cite{wang2026tars} uses reinforcement learning with representation and behavior alignment rewards. TextPro-SLM~\cite{cui2026textpro} provides transcribed text and prosodic features through its WhisperPro encoder. These approaches address input-side capability gaps; we instead examine output-mode differences with the speech input and model fixed.

\subsection{Cross-modal on-policy distillation}

On-policy distillation transfers teacher capabilities through supervision on student-generated trajectories~\cite{agarwal2024opd}. X-OPD~\cite{cao2026xopd} uses a text-conditioned teacher to supervise student-generated trajectories under both speech and text inputs. CORD~\cite{hu2026cord} uses an internal text-conditioned teacher with weighted token-level distillation and sequence-level reinforcement learning. $X^3$-OPD~\cite{fu2026x3opd} extends supervision to audio-grounded reasoning using matched text inputs and reference answers. JO-OPD instead keeps the speech input unchanged and provides S2T supervision on text-only projections of student-generated S2TS histories.

\section{Method}

\subsection{Output-mode gap}

We consider speech LLMs supporting both speech-to-text (S2T) and speech-to-text-and-speech (S2TS) generation. Given speech input $x^S$, S2T produces a text response, whereas S2TS produces a joint trajectory $z$ containing text tokens, acoustic outputs, and control tokens. Let $P_T$ denote the text-only projection that retains text tokens in their original order and removes acoustic outputs and control tokens. We denote the internal text $P_T(z)$ by S2TS(T) and the reconstructed spoken response by S2TS(S).

Let $\mathcal{A}$ denote answer accuracy (\%) on a shared set of speech inputs $X^S$. We define the output-mode gap (OMG) as
\begin{equation}
\begin{aligned}
\Delta_{\mathrm{OMG}}
&= \mathcal{A}_{\mathrm{S2T}}
 - \mathcal{A}_{\mathrm{S2TS(T)}} \\
&= \mathcal{A}\big(\mathrm{S2T}(X^S)\big)
 - \mathcal{A}\big(P_T(\mathrm{S2TS}(X^S))\big).
\end{aligned}
\label{eq:gap}
\end{equation}
Both accuracies are computed using the same model, evaluation examples, and answer-scoring rule. A positive OMG indicates lower internal-text answer accuracy when speech output is enabled. Spoken-answer accuracy is assessed separately by transcribing S2TS(S) and scoring the extracted answer.

\subsection{Joint-Output On-Policy Distillation}

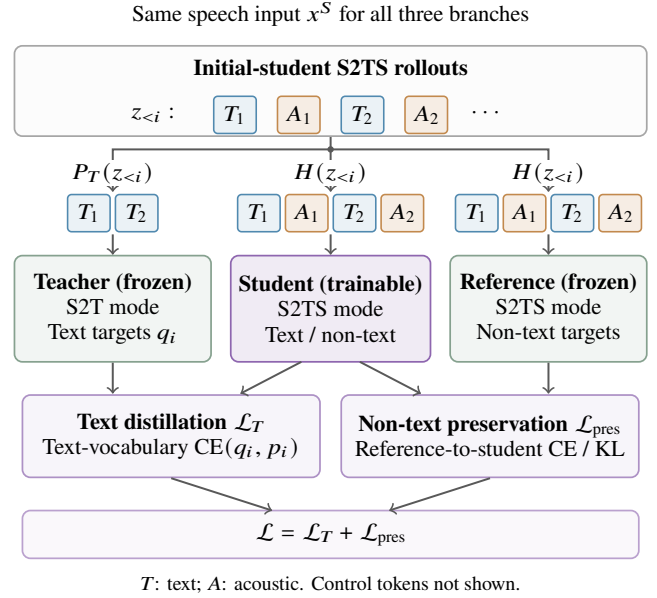
\begin{figure}[t]
  \centering
\begingroup
\definecolor{joText}{HTML}{246B9B}
\definecolor{joAudio}{HTML}{B36B20}
\definecolor{joFrozen}{HTML}{5D826A}
\definecolor{joTrainable}{HTML}{8155A6}
\begin{tikzpicture}[x=1pt,y=1pt,
  every node/.style={font=\fontsize{8.5}{10}\selectfont,align=center,inner sep=3pt},
  panel/.style={draw=black!35,rounded corners=3pt,line width=.65pt},
  flow/.style={->,line width=.75pt,draw=black!65},
  token/.style={rounded corners=1.5pt,minimum width=16pt,
    minimum height=14pt,inner sep=1pt,line width=.6pt},
  texttoken/.style={token,draw=joText!80,fill=joText!9},
  audiotoken/.style={token,draw=joAudio!80,fill=joAudio!12},
  context/.style={fill=white,inner sep=1pt}]

  \node at (122,231) {Same speech input $x^S$ for all three branches};
  \node[panel,fill=black!1,minimum width=238pt,minimum height=34pt]
    (rollout) at (122,202) {};
  \node[font=\fontsize{8.5}{10}\selectfont\bfseries] at (122,211)
    {Initial-student S2TS rollouts};
  \node at (55,193) {$z_{<i}:$};
  \node[texttoken] at (86,193) {$T_1$};
  \node[audiotoken] at (110,193) {$A_1$};
  \node[texttoken] at (134,193) {$T_2$};
  \node[audiotoken] at (158,193) {$A_2$};
  \node at (181,193) {$\cdots$};

  \draw[flow] (rollout.south) -- (122,180) -- (40,180) -- (40,164);
  \draw[flow] (122,180) -- (122,164);
  \draw[flow] (122,180) -- (204,180) -- (204,164);
  \fill[black!65] (122,180) circle (1.2pt);
  \node[context] at (40,171) {$P_T(z_{<i})$};
  \node[context] at (122,171) {$H(z_{<i})$};
  \node[context] at (204,171) {$H(z_{<i})$};

  \node[texttoken] at (31,156) {$T_1$};
  \node[texttoken] at (49,156) {$T_2$};
  \foreach \offset in {122,204} {
    \node[texttoken] at (\offset-27,156) {$T_1$};
    \node[audiotoken] at (\offset-9,156) {$A_1$};
    \node[texttoken] at (\offset+9,156) {$T_2$};
    \node[audiotoken] at (\offset+27,156) {$A_2$};
  }

  \node[panel,draw=joFrozen!75,fill=joFrozen!8,
    minimum width=74pt,minimum height=40pt]
    (teacher) at (40,120)
    {\textbf{Teacher (frozen)}\\S2T mode\\Text targets $q_i$};
  \node[panel,draw=joTrainable!90,fill=joTrainable!12,
    minimum width=74pt,minimum height=40pt]
    (student) at (122,120)
    {\textbf{Student (trainable)}\\S2TS mode\\Text / non-text};
  \node[panel,draw=joFrozen!75,fill=joFrozen!8,
    minimum width=74pt,minimum height=40pt]
    (reference) at (204,120)
    {\textbf{Reference (frozen)}\\S2TS mode\\Non-text targets};
  \draw[flow] (40,148) -- (teacher.north);
  \draw[flow] (122,148) -- (student.north);
  \draw[flow] (204,148) -- (reference.north);

  \node[panel,draw=joTrainable!65,fill=joTrainable!6,
    minimum width=112pt,minimum height=31pt]
    (textloss) at (62,72)
    {\textbf{Text distillation $\mathcal L_T$}\\
     Text-vocabulary $\operatorname{CE}(q_i,p_i)$};
  \node[panel,draw=joTrainable!65,fill=joTrainable!6,
    minimum width=112pt,minimum height=31pt]
    (preserve) at (182,72)
    {\textbf{Non-text preservation $\mathcal L_{\mathrm{pres}}$}\\
     Reference-to-student CE / KL};
  \draw[flow] (teacher.south) -- (40,88);
  \draw[flow] (111,100) -- (88,88);
  \draw[flow] (133,100) -- (156,88);
  \draw[flow] (reference.south) -- (204,88);

  \node[panel,draw=joTrainable!55,fill=joTrainable!4,
    minimum width=232pt,minimum height=18pt]
    (objective) at (122,35)
    {$\mathcal L=\mathcal L_T+\mathcal L_{\mathrm{pres}}$};
  \draw[flow] (textloss.south) -- (100,44);
  \draw[flow] (preserve.south) -- (144,44);
  \node[font=\fontsize{7.5}{9}\selectfont,inner sep=1pt] at (122,16)
    {$T$: text; $A$: acoustic. Control tokens not shown.};
\end{tikzpicture}
\endgroup
  \caption{Overview of JO-OPD. Teacher and reference share a frozen base model, using S2T on projected text histories and S2TS on joint histories, respectively.}
  \label{fig:method}
\end{figure}

As shown in Fig.~\ref{fig:method}, JO-OPD uses the model's stronger S2T policy to supervise text predictions along student-generated S2TS trajectories. The base model serves two roles: its S2T policy provides text supervision, and its S2TS policy provides reference distributions for preserving non-text predictions.

\noindent\textbf{History projection.} The student is initialized from the base model. Given speech input $x^S$, we sample a joint trajectory $z$ from the student in S2TS mode. At each text prediction position $i$, we construct two contexts:
\begin{align}
  h_i^{\mathrm{stu}}
  &= \bigl(x^S,H(z_{<i}),\mathrm{S2TS}\bigr), \notag\\
  h_i^{\mathrm{tea}}
  &= \bigl(x^S,P_T(z_{<i}),\mathrm{S2T}\bigr).
  \label{eq:contexts}
\end{align}
Here, $H(z_{<i})$ denotes the student's native joint output history at position $i$. The teacher receives the same speech input and the student's generated text prefix, while the student also conditions on its acoustic history.

\noindent\textbf{Text distillation.} Text distillation compares teacher and student predictions within the text vocabulary. This supervises relative preferences among text tokens without directly matching the probability assigned to text versus non-text outputs.

Let $p_i$ denote the student distribution under $h_i^{\mathrm{stu}}$ and $q_i$ the teacher target under $h_i^{\mathrm{tea}}$. The student distribution is normalized over the full text vocabulary. JO-OPD-Soft obtains $q_i$ by retaining the teacher's top-$k$ text tokens and renormalizing their probabilities. JO-OPD-Hard instead uses a one-hot target at the teacher's highest-probability text token. Both variants minimize the same text distillation objective:
\begin{equation}
\mathcal L_T
= \mathbb E_{z,i}\big[\operatorname{CE}(q_i,p_i)\big],
\label{eq:loss}
\end{equation}
where $\operatorname{CE}$ denotes cross-entropy, and the expectation averages over training trajectories and sampled text positions within each trajectory.

\noindent\textbf{Non-text preservation.} Updates to shared parameters may also change predictions of acoustic outputs and control tokens. We therefore regularize these predictions against the fixed base model in S2TS mode, with both models conditioned on the same native joint history.

For Step-Audio~2, we apply cross-entropy at sampled non-text positions of the shared output head, using renormalized top-$k$ reference targets selected from the full output vocabulary. Student probabilities are normalized over the full vocabulary. For Baichuan-Audio, we separately regularize the controller and codec distributions using exact reference-to-student KL divergence. Gradients from the codec loss propagate through the frozen speech head to the shared language model.

Let $\mathcal L_{\mathrm{pres}}$ denote the weighted sum of mean distribution-matching losses for these non-text predictions. The overall objective is
\begin{equation}
\mathcal L
= \mathcal L_T + \mathcal L_{\mathrm{pres}}.
\label{eq:total-loss}
\end{equation}

\section{Experimental Setup}

\subsection{Models and datasets}

We evaluate JO-OPD on Step-Audio-2-mini and Baichuan-Audio-Instruct. The training set comprises 27,847 prompts from Tulu~3~\cite{lambert2024tulu3} and NaturalReasoning~\cite{yuan2025naturalreasoning}, excluding overlap with the evaluation benchmarks. Training speech is synthesized using the SLT voice in flite. We evaluate answer accuracy on Spoken-MQA~\cite{wei2025spokenmqa} (1,402 questions) and speech-rendered GSM8K~\cite{cobbe2021gsm8k} (1,319 test questions). To examine the effect of reasoning instructions beyond mathematical tasks, we additionally evaluate Step-Audio~2 on 1,000 human-recorded questions from four VoiceBench-BBH~\cite{chen2024voicebench} tasks under short-answer and explicit-reasoning conditions.

\subsection{Training and comparisons}

In the main experiments, we train JO-OPD for one epoch on fixed S2TS trajectories sampled from the initial student. For Step-Audio~2, we update the full language model while freezing the audio encoder and adapter. For Baichuan-Audio, we update the full language model and text head, with all other components frozen. The teacher and preservation reference remain fixed throughout training. We use AdamW with a batch size of 32 and a learning rate of $2\times10^{-6}$.

JO-OPD-Soft uses top-32 teacher targets for text distillation. For Step-Audio~2, the preservation loss uses top-32 reference targets with a weight of 0.05. For Baichuan-Audio, the controller and codec preservation losses are weighted by 1 and 10, respectively.

We compare JO-OPD with Self-SFT and Response-SFT on Step-Audio~2, and with Response-SFT on Baichuan-Audio. Self-SFT uses the same S2TS trajectories, while Response-SFT uses independently generated S2T responses. Each baseline matches JO-OPD's training prompts and optimization updates on its backbone.

\subsection{Decoding and evaluation}

Evaluation follows each model's native decoding policy, with matched inputs and generation seeds for base--student comparisons. We generate one response per input and output mode. The generation budget is 4,096 joint tokens for Step-Audio~2 and 4,096 text tokens for Baichuan-Audio.

For numerical benchmarks, we report answer accuracy using the same extractor across models and output modes. Spoken responses are transcribed using Whisper-large-v3-turbo and evaluated with the same scoring rule. BBH uses task-specific extraction of yes/no responses or option labels. All differences are computed from unrounded accuracies.

\section{Results and Analysis}

\subsection{Reducing the output-mode gap}
\label{sec:cross-model}

As shown in Table~\ref{tab:main}, JO-OPD-Soft substantially reduces OMG in Step-Audio~2. The gap decreases from 42.87 to 16.26 points on Spoken-MQA and from 29.72 to 13.04 points on GSM8K. Since S2T accuracy remains nearly unchanged, these reductions primarily reflect improved text generation in S2TS mode.

The matched baselines provide further insight into the supervision strategy. Self-SFT decreases S2TS(T) accuracy on both benchmarks, while Response-SFT offers little improvement in this mode. JO-OPD uses the stronger S2T policy to provide feedback at the student's own joint-generation prefixes. Its larger gains suggest that aligning supervision with these prefixes is useful for improving text predictions conditioned on acoustic history.

\begin{table}[t]
\centering
\caption{Step-Audio~2 text answer accuracy (\%) and OMG (pp).}
\label{tab:main}
\normalsize
\setlength{\tabcolsep}{1.0pt}
\renewcommand{\arraystretch}{1.08}
\begin{tabular*}{\columnwidth}{@{\extracolsep{\fill}}lcccccc@{}}
\toprule
& \multicolumn{3}{c}{Spoken-MQA} & \multicolumn{3}{c}{GSM8K}\\
\cmidrule(lr){2-4}\cmidrule(lr){5-7}
Method & S2T & S2TS(T) & OMG $\downarrow$ & S2T & S2TS(T) & OMG $\downarrow$\\
\midrule
Base & 75.32 & 32.45 & 42.87 & 69.22 & 39.50 & 29.72\\
Self-SFT & 76.18 & 23.61 & 52.57 & 71.72 & 36.92 & 34.80\\
Response-SFT & 75.96 & 31.60 & 44.37 & 71.27 & 42.76 & 28.51\\
JO-OPD-Hard & \textbf{77.03} & 40.94 & 36.09 & \textbf{75.13} & 52.54 & 22.59\\
\textbf{JO-OPD-Soft} & 75.18 & \textbf{58.92} & \textbf{16.26} & 69.90 & \textbf{56.86} & \textbf{13.04}\\
\bottomrule
\end{tabular*}
\end{table}

Table~\ref{tab:baichuan} shows that JO-OPD also reduces OMG in Baichuan-Audio, by 2.92 points on Spoken-MQA and 0.61 points on GSM8K. Here, S2T accuracy also improves, making the reduction in OMG smaller than the gain in S2TS(T) accuracy. JO-OPD also exceeds Response-SFT in S2TS(T) accuracy on both benchmarks, with a larger gain on Spoken-MQA.

\begin{table}[t]
\centering
\caption{Baichuan-Audio text answer accuracy (\%) and OMG (pp).}
\label{tab:baichuan}
\normalsize
\setlength{\tabcolsep}{1.5pt}
\renewcommand{\arraystretch}{1.12}
\begin{tabular*}{\columnwidth}{@{\extracolsep{\fill}}llccc@{}}
\toprule
Dataset & Method & S2T & S2TS(T) & OMG $\downarrow$\\
\midrule
\multirow{3}{*}{Spoken-MQA}
& Base & 66.26 & 53.85 & 12.41\\
& Response-SFT & 66.12 & 55.56 & 10.56\\
& JO-OPD & \textbf{69.54} & \textbf{60.06} & \textbf{9.49}\\
\midrule
\multirow{3}{*}{GSM8K}
& Base & 58.23 & 47.84 & 10.39\\
& Response-SFT & \textbf{59.36} & 46.55 & 12.81\\
& JO-OPD & 58.76 & \textbf{48.98} & \textbf{9.78}\\
\bottomrule
\end{tabular*}
\end{table}

The results in Table~\ref{tab:voicebench-bbh} further show that explicit reasoning instructions affect the two output modes differently. On human-recorded VoiceBench-BBH questions, requesting explicit reasoning slightly increases Step-Audio~2's Base S2T accuracy but decreases S2TS(T) accuracy, widening OMG from 8.2 to 19.0 points. JO-OPD-Soft improves S2TS(T) accuracy by 14.7 points under CoT, compared with 2.9 points under Short, leaving gaps of 5.8 and 5.0 points, respectively. The larger CoT gain suggests that JO-OPD is particularly useful when reasoning instructions expose a greater discrepancy between output modes. These results extend its benefits to human-recorded reasoning questions beyond the numerical benchmarks.

\begin{table}[t]
\centering
\caption{Step-Audio~2 answer accuracy (\%) and OMG (pp) under Short and CoT conditions on the same 1,000 human-recorded VoiceBench-BBH inputs.}
\label{tab:voicebench-bbh}
\normalsize
\setlength{\tabcolsep}{2pt}
\renewcommand{\arraystretch}{1.08}
\begin{tabular*}{\columnwidth}{@{\extracolsep{\fill}}llccc@{}}
\toprule
Condition & Method & S2T & S2TS(T) & OMG $\downarrow$\\
\midrule
\multirow{3}{*}{Short}
& Base & \textbf{56.8} & 48.6 & 8.2\\
& Response-SFT & 56.3 & 50.4 & 5.9\\
& JO-OPD-Soft & 56.5 & \textbf{51.5} & \textbf{5.0}\\
\midrule
\multirow{3}{*}{CoT}
& Base & 58.6 & 39.6 & 19.0\\
& Response-SFT & 59.5 & 43.1 & 16.4\\
& JO-OPD-Soft & \textbf{60.1} & \textbf{54.3} & \textbf{5.8}\\
\bottomrule
\end{tabular*}
\end{table}

\subsection{Spoken-answer accuracy}

As shown in Table~\ref{tab:speech-realization}, JO-OPD improves spoken-answer accuracy on Spoken-MQA for both models. Step-Audio~2's S2TS(S) accuracy increases from 28.67\% to 36.16\%, a gain of 7.49 points, while Baichuan-Audio improves by 2.92 points. Baichuan-Audio also gains 2.35 points on GSM8K, where Step-Audio~2 shows little change from Base.

On Spoken-MQA, JO-OPD-Soft improves Step-Audio~2's spoken-answer accuracy over Response-SFT by 7.85 points. For Baichuan-Audio, the spoken-answer gains over Response-SFT are modest, at approximately 1.5 percentage points on both benchmarks.

\begin{table}[t]
\centering
\caption{Internal-text and ASR-based spoken-answer accuracy (\%) in S2TS mode.}
\label{tab:speech-realization}
\normalsize
\setlength{\tabcolsep}{1.5pt}
\renewcommand{\arraystretch}{1.12}
\begin{tabular*}{\columnwidth}{@{\extracolsep{\fill}}llcccc@{}}
\toprule
& & \multicolumn{2}{c}{Spoken-MQA} & \multicolumn{2}{c}{GSM8K}\\
\cmidrule(lr){3-4}\cmidrule(lr){5-6}
Model & Method & \shortstack{S2TS\\(T)} & \shortstack{S2TS\\(S)}
& \shortstack{S2TS\\(T)} & \shortstack{S2TS\\(S)}\\
\midrule
\multirow{3}{*}{Step-Audio-2}
& Base & 32.45 & 28.67 & 39.50 & 36.16\\
& Response-SFT & 31.60 & 28.32 & 42.76 & \textbf{39.80}\\
& JO-OPD-Soft & \textbf{58.92} & \textbf{36.16} & \textbf{56.86} & 35.25\\
\midrule
\multirow{3}{*}{Baichuan-Audio}
& Base & 53.85 & 50.14 & 47.84 & 43.82\\
& Response-SFT & 55.56 & 51.57 & 46.55 & 44.66\\
& JO-OPD & \textbf{60.06} & \textbf{53.07} & \textbf{48.98} & \textbf{46.17}\\
\bottomrule
\end{tabular*}
\end{table}

\subsection{Ablation studies}

As shown in Table~\ref{tab:main}, JO-OPD-Soft exceeds JO-OPD-Hard in S2TS(T) accuracy by 17.97 points on Spoken-MQA and 4.32 points on GSM8K. With the teacher, trajectories, and history projection held fixed, this comparison favors retaining the teacher's relative preferences among candidate text tokens over using only its highest-probability prediction.

Table~\ref{tab:ablations} examines text-vocabulary restriction and non-text preservation. Both ablations yield internal-text accuracies within 1.8 points of JO-OPD-Soft, but substantially lower spoken-answer accuracy. On Spoken-MQA, full-vocabulary supervision reduces S2TS(S) accuracy from 36.16\% to 11.20\%, while removing preservation reduces it to 29.74\%. The same pattern holds on GSM8K. These results support restricting text supervision to the text vocabulary while separately regularizing non-text predictions to preserve speech generation.

\begin{table}[t]
\centering
\caption{Ablations of text-vocabulary restriction and non-text preservation on Step-Audio~2. We report internal-text and ASR-based spoken-answer accuracy (\%).}
\label{tab:ablations}
\normalsize
\setlength{\tabcolsep}{1.5pt}
\renewcommand{\arraystretch}{1.12}
\begin{tabular*}{\columnwidth}{@{\extracolsep{\fill}}lcccc@{}}
\toprule
& \multicolumn{2}{c}{Spoken-MQA} & \multicolumn{2}{c}{GSM8K}\\
\cmidrule(lr){2-3}\cmidrule(lr){4-5}
Method & S2TS(T) & S2TS(S) & S2TS(T) & S2TS(S)\\
\midrule
JO-OPD-Soft & 58.92 & \textbf{36.16} & 56.86 & \textbf{35.25}\\
Full vocabulary & \textbf{60.70} & 11.20 & 55.80 & 13.04\\
No preservation & 59.99 & 29.74 & \textbf{58.38} & 30.02\\
\bottomrule
\end{tabular*}
\end{table}

\subsection{Training scale and trajectory renewal}

We evaluate JO-OPD-Soft on Step-Audio~2 using nested subsets of the same initial-student trajectory pool. Table~\ref{tab:training-scale} shows that S2TS(T) accuracy increases with training set size, while S2T accuracy remains near 75\%. Under one-epoch training, larger datasets also entail more optimization updates, so this trend reflects the combined effects of additional data and optimization.

\begin{table}[t]
\centering
\caption{Step-Audio~2 text answer accuracy (\%) on Spoken-MQA with different JO-OPD-Soft training set sizes.}
\label{tab:training-scale}
\normalsize
\setlength{\tabcolsep}{1.5pt}
\renewcommand{\arraystretch}{1.12}
\begin{tabular*}{\columnwidth}{@{\extracolsep{\fill}}lcc@{}}
\toprule
Training prompts & S2T & S2TS(T)\\
\midrule
0 (Base) & 75.32 & 32.45\\
2,000 & \textbf{75.89} & 36.95\\
8,000 & 75.04 & 49.71\\
27,847 & 75.18 & \textbf{58.92}\\
\bottomrule
\end{tabular*}
\end{table}

For trajectory renewal, we compare Static, which retains initial-student trajectories, with Refresh, which regenerates trajectories from the updated student before stages 2--4. Both four-stage runs use the same prompt partitions and 871 updates, with the optimizer and learning-rate schedule reset at each stage. Refresh improves S2TS(T) accuracy by 2.57 points on Spoken-MQA and 2.73 points on GSM8K, with little change in S2T accuracy.

\section{Conclusion}

We identify an output-mode gap in interleaved speech LLMs, where enabling speech output can reduce text answer accuracy. To reduce this gap, we propose JO-OPD, which uses the model's stronger S2T policy to supervise student-generated S2TS trajectories through text-only history projection while regularizing native non-text predictions. Experiments on Step-Audio~2 and Baichuan-Audio show OMG reductions in both models, with JO-OPD outperforming matched SFT baselines on Step-Audio~2. Ablations support the use of soft text targets, vocabulary restriction, and non-text preservation. The observed spoken-answer gains on Spoken-MQA further suggest that supervision from the model's own S2T policy can benefit the content of spoken responses.

\clearpage
\bibliographystyle{IEEEbib}
\bibliography{references}

\end{document}